\documentclass[11 pt]{article}
\usepackage{graphicx}
\begin{document}

\title{Shape and Angular Distribution of the 4.438-MeV Line from Proton Inelastic Scattering off $^{12}$C }

\author{J. Kiener \\ 
Centre de Sciences Nucl\'eaires et de Sciences de la Mati\`ere (CSNSM), \\ Univ. Paris-Sud, CNRS/IN2P3, Universit\'e Paris-Saclay, 91405 Orsay, France}

\maketitle

\begin{abstract}
The emission of the 4.438-MeV $\gamma$-ray line in proton inelastic scattering   off $^{12}$C has been investigated in detail. The correlated scattering and emission process is described independently for the direct reaction mechanism and for the compound-nucleus (CN) component.  The inelastic scattering process  for direct reactions is treated  with a coupled-channels nuclear reaction code, while the CN component is described as a superposition of separate resonances with definite spin and parity, treated with the angular momentum coupling theory. The calculations are compared to a comprehensive data set on measured line shapes and angular distributions in the proton energy range $E_p$ = 5.44 - 25.0 MeV.  In the range  $E_p$ $\approx$ 12 - 25 MeV a good agreement is obtained in calculations assuming direct reactions with only a negligible part of CN reactions. At lower energy, the data are reproduced by incoherent sums of the direct component with typically one CN resonance. Based on these results, the prospectives for line shape calculations applied to solar flares and $\gamma$-ray emission in proton radiotherapy are discussed. 
\end{abstract}

\section{Introduction}

The shape of the 4.438-MeV line of $^{12}$C emitted during solar flares is an important observable that may be used to determine the composition, energy spectra and directional distribution of the accelerated particles  interacting in the solar atmosphere. Up to now, its line shape could be studied in several flares observed by SMM \cite{SMKS}, Rhessi \cite{Lin03} and INTEGRAL  \cite{Ocflare,Harris}. In particular observations with the high-energy resolution Ge detectors onboard the Rhessi and INTEGRAL satellites permitted fine spectroscopy of the line and resulted in interesting conclusions about the accelerated proton and $\alpha$-particle directional distributions in the impulsive flare phases.

Emission of this line from the inner Galaxy due to interactions of low-energy cosmic rays, is also one of the primary science objectifs of next-generation $\gamma$-ray observatories \cite{Galcenter,eAstrogam}. A good knowledge of the line shape there should facilitate its extraction in the presence of multiple other emissions and the determination of its intensity. In astrophysical sites in general, the line is mainly produced by energetic-proton and $\alpha$-particle induced reactions with $^{12}$C and $^{16}$O of the interaction medium and in inverse kinematics by reactions  of accelerated $^{12}$C and $^{16}$O with helium and hydrogen. 

Furthermore, there is significant interest in the monitoring of the dose deposition in radiotherapy with particle beams \cite{DP16}. An example are devices that detect the prompt $\gamma$ rays resulting from proton (or carbon) interactions in human tissue, rising the demand of detailed and reliable nuclear cross sections \cite{Verburg12,Jeya15}. In this context, carbon and oxygen make up more than 3/4 of the human body mass, making the 4.438-MeV line the strongest prompt emission line produced.

In all described scenarios, inelastic scattering of protons off $^{12}$C makes up a significant part of the 4.438-MeV line intensity.  The line-shape calculations for this component in the solar flare studies \cite{SMKS,Ocflare,Harris} were based on a relatively simple parametrization of the magnetic substate population of the 2$^+$ 4.439-MeV state of $^{12}$C at low proton energies $E_p$ $<\sim$ 15 MeV, and coupled-channels calculations at higher energies adjusted to reproduce an abundant data set of measured line shapes and $\gamma$-ray angular distributions \cite{lshape}. This parametrization reproduced fairly well the meaured line shapes and $\gamma$-ray angular distributions above $E_p$ = 15 MeV, but reproduced only approximately the experimental data at lower proton energies.

\begin{figure}[htb]
\centering
\includegraphics[scale=0.5]{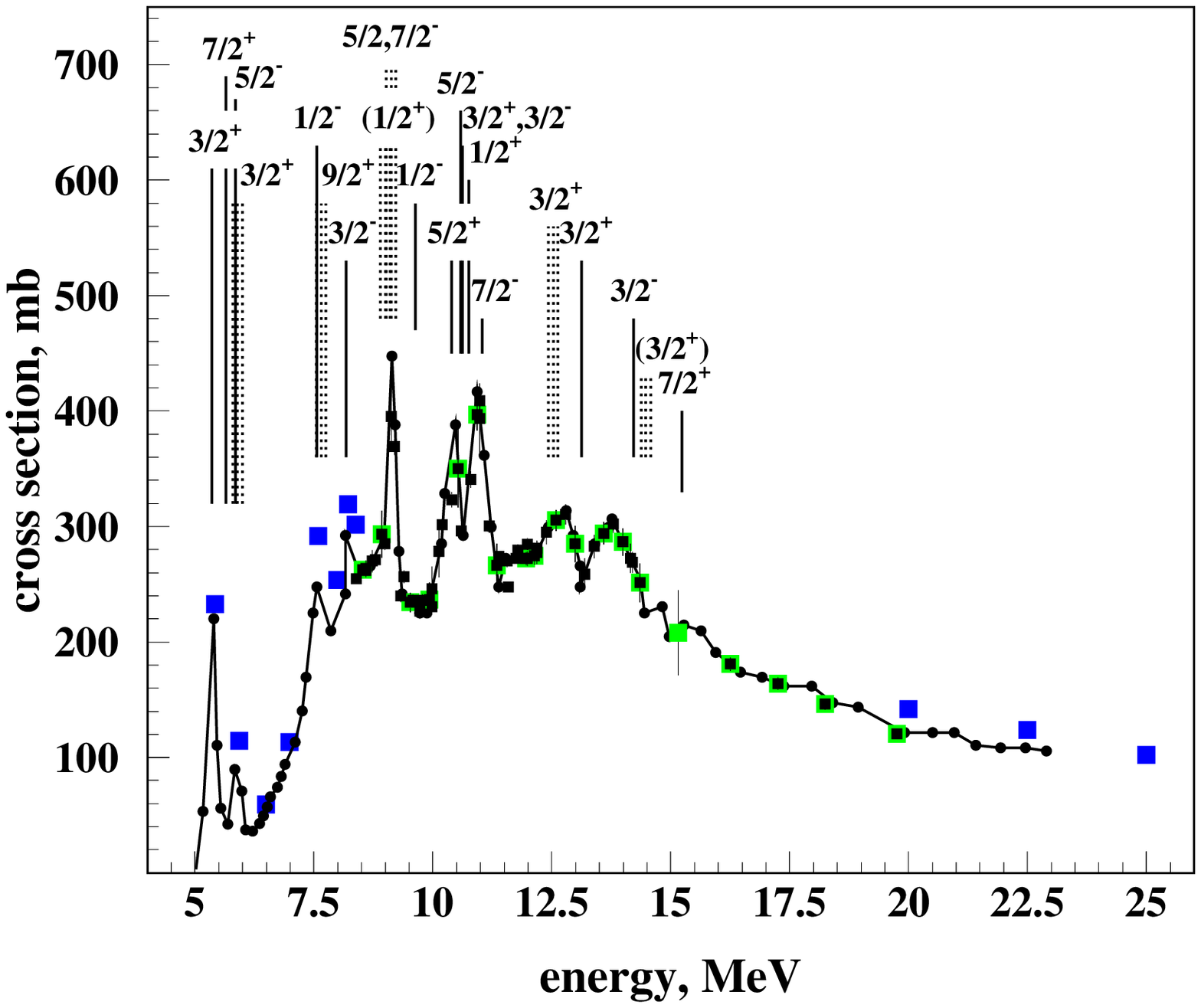}
\caption[]{Cross section as a fonction of laboratory proton energy for emission of the 4.438-MeV $\gamma$ ray in proton inelastic scattering off $^{12}$C. Data are from Dyer et al. \cite{Dyer81} (black dots), the Orsay-1997 experiment \cite{torion} (black squares) and the Orsay-2002 experiment \cite{t2002} (blue squares).  Larger green squares indicate Orsay-1997 data where line shapes are available. Vertical lines indicate the position of  $^{13}$N states in the range $E_x$ = 6.5 - 18 MeV with spin and parity assignment in the NNDC data base \cite{NNDC}. }
\label{Excf44}
\end{figure}

Since then, new  data for proton inelastic scattering off $^{12}$C became available, resulting in a now full data set of line shapes and angular distributions for the 4.438-MeV line in the proton energy range from threshold to $E_p$ = 25 MeV. There are now, in particular, line-shape data close to the energy of each CN resonance that shows up as a distinct peak in the  $\gamma$-ray production cross section in proton reactions with $^{12}$C (see Figure \ref{Excf44}).  With these data, a new method has been developed that  improves specifically the description of line shapes and $\gamma$-ray angular distributions in the region of dominating CN resonances, i.e. from threshold to about $E_p$ = 12 MeV.  At higher proton energies, coupled-channel calculations with a deformed  potential for nucleon scattering off $^{12}$C \cite{Meigooni85}  reproduced fairly well the measured data. 

In the following section, a short outline of the formalism for the calculations will be given. After that, the extended parameter search for these calculations to reproduce measured $\gamma$-ray line shapes and angular distributions is described and the results are discussed. 

\section{Calculations}

The 4.438-MeV line emission results from the deexcitation of the first excited state of $^{12}$C, 2$^+$ at 4.439 MeV with a halflife of 61 fs. This level belongs to the ground-state rotational band and is thus strongly coupled to the 0$^+$ ground state, which results in relatively high direct reaction cross sections in proton inelastic scattering off $^{12}$C. The second excited state of $^{12}$C, a 0$^+$ state at 7.654 MeV, is already above the $\alpha$-particle emission threshold and has only a very weak  $\gamma$-decay branching,  as generally  for all higher-excited states. Population of the 4.439 MeV-state by cascades from  higher-lying levels can therefore be neglected, which facilitates greatly the line-shape calculations. A possible contribution from the 4.445-MeV state of $^{11}$B, populated by the $^{12}$C(p,2p)$^{11}$B reaction was found to be weak in the explored energy range.

The cross section for emission of the 4.438-MeV line in p + $^{12}$C reactions reveals several distinct peaks in the proton energy range below $E_p$ $\sim$15 MeV (see Figure \ref{Excf44}), probably due to isolated CN resonances. It was therefore decided to reproduce the measured line shapes and $\gamma$-ray angular distributions by incoherent sums of the direct reaction component and a CN component corresponding to a resonance state in $^{13}$N with definite spin and parity $J^{\pi}$. 

\subsection{Direct reaction component} 

Proton inelastic scattering to states of a collective band like the 4.439-MeV state of $^{12}$C is usually  satisfactorily described by nuclear reaction calculations in the coupled-channels framework (see e.g. \cite{Satchler} and references therein). This is readily done with nuclear reaction codes like Ecis \cite{Ecis}, that provide the necessary flexibility in the reaction parameter input to define the ground and excited states and their couplings, potentials etc. The angular correlation between the scattered proton and the emitted $\gamma$ ray $W(\theta_p,\theta_{\gamma},\phi_{\gamma})$ is calculated with the scattering amplitudes, that may be obtained directly from the nuclear reaction code or calculated from the scattering matrix elements as e.g. given in the output of the code Ecis. The formalism for these calculations is broadly outlined in Ref. \cite{lshape} and detailed in Ref. \cite{newShape}.

Such calculations have already been shown to reproduce  measured 4.438-MeV line shapes and $\gamma$-ray angular distributions at proton energies where the compound-nucleus component is negligible, \i.e. above $E_p$ = 15 MeV \cite{Werntz90,lshape}. In both references, the coupled-channel calculations were done with rotational coupling of only two states, the $^{12}$C ground state and the 4.439-MeV, 2$^+$ state.
The same simple approach was adopted here, using  the nuclear reaction potential of Meigooni et al. \cite{Meigooni85} above $E_p$ = 15 MeV. At lower proton energies, other potentials from the compilation of Perey\&Perey \cite{Perey} were also tried. 

\subsection{Compound nucleus component} 

At each proton energy, where line-shape data are available, calculations for all possible CN resonance states in $^{13}$N were done. Properties of $^{13}$N states were taken from the NNDC data base \cite{NNDC}.  Inelastic scattering in the CN formulism proceeds through the formation of an intermediate CN resonance of $J^{\pi}$ = $b$, that decays to the excited state of the target with $J^{\pi}$ = $c$ by particle emission with angular momentum $L$ followed by  a deexcitation $\gamma$ ray with $L_{\gamma}$  to the target state $d$. The formation of a CN resonance with definite spin and parity $b$ in the p + $^{12}$C reaction ($J^{\pi}$ = 0$^+$ and $s$ = 1/2$^+$ for $^{12}$C$_{g.s.}$ and proton spin, respectively) results in a unique incoming angular momentum for $l_i$ from angular momentum selection rules.  The subsequent steps:

\begin{equation} (1) ~~~  \vec{b} = \vec{c} + \vec{L}  ~~~~~ (2)~~~  \vec{c} = \vec{d} + \vec{L_{\gamma}}
\label{bcL}
\end{equation}

with $c$ = 2$^+$ and $d$ = 0$^+$ for the $^{12}$C excited and ground states, respectively, leaves a unique $L_{\gamma}$, while the orbital angular momentum $l$ ($\vec{L}$ = $\vec{l}$ + $\vec{s}$) of the proton emitted by the CN state can take  different values. The formulism for the angular correlation of outgoing proton and the $\gamma$ ray  $W(\theta_p,\theta_{\gamma},\phi_{\gamma})$ has been taken from the monograph of Ferguson \cite{Ferguson} and is detailed in \cite{newShape}. 

Practically, all measured data could be reasonably reproduced with the 2 smallest orbital angular momenta $l$ of the outgoing proton, the only free parameter in the calculation of   $W(\theta_p,\theta_{\gamma},\phi_{\gamma})$ for a given resonance is then the branching ratio $W_{l_0}$

\begin{equation} 
W_{l_0} := \frac{W(l_0)}{W(l_0) + W(l_0 + 2)} 
\label{W02}
\end{equation} 

where $W(l_0)$ is the probability for proton emission with the lowest possible orbital angular momentum $l_0$. The rest of the calculation is fixed by the angular momentum coupling rules.

\subsection{Simulation of experimental data}

Line-shape calculations for the CN and direct components were realized with two separate Monte-Carlo type programs. Both are nearly indentical except for the formalism describing the angular correlations,  consisting in event-by-event simulation of the reaction sequence from the inelastic scattering reaction to the emission and detection of the $\gamma$ ray. The interaction depth in the target, center-of-mass scattering angle $\theta_p$, time of the $\gamma$-ray emission and $\gamma$-ray emission angles $\theta_{\gamma},\phi_{\gamma}$ are randomly drawn from cumulative distributions reflecting the different probabilities. For the correlated proton and $\gamma$-ray emission, the correlation functions $W(\theta_p, \theta_{\gamma},\phi_{\gamma})$ are represented by a series of cumulative ($\theta_{\gamma},\phi_{\gamma}$)-distributions  for a grid of scattering angles with $\Delta \theta_p$ = 5$^{\circ}$ in the case of CN resonances and $\Delta \theta_p$ = 10$^{\circ}$ for direct reactions.

Recoil angle and energy of the excited $^{12}$C in the laboratory are determined from nuclear reaction kinematics and the slowing down of the recoil $^{12}$C in the target is taken into account until deexcitation of the 4.439-MeV state occurs, the recoil nucleus stops or leaves the target. Polar angle and energy of the emitted $\gamma$ ray in the laboratory are calculated with relativistic kinematics and stored in an energy-angle histogram.  Line shapes are then constructed from the energy-angle histogram for  the solid angles spanned by the detectors used in the different experiments, the effect of detector energy resolution on the line shapes being included. A few million events per proton energy were typically simulated for each component, resulting in high-statistics calculated line profiles and laboratory $\gamma$-ray angular distributions.

\section{Results}

The bulk of available line shape data for the 4.438-MeV line in proton inelastic scattering off $^{12}$C are from two experiments done at the Orsay tandem Van-de-Graaff accelerator in 1997 \cite{torion} and 2002 \cite{t2002}. They cover the proton energy range from threshold to $E_p$ = 25 MeV with line shapes existing for a total of 31 different proton energies. In both Orsay experiments, $\gamma$-ray spectra have been obtained with large-volume coaxial HP-Ge detectors equipped with active BGO shielding. In the 1997 experiment, 8 detectors were used in the angular range $\theta$ = 45$^{\circ}$ - 145$^{\circ}$, and 5 or 4 detectors in the 2002 experiment in the range $\theta$ = 45$^{\circ}$ or 67.5$^{\circ}$ - 157.5$^{\circ}$. There are only two other data sets with line shapes at $E_p$ = 23 MeV \cite{Kolata67} and at $E_p$ = 40 MeV \cite{Lang87}.

In the region below $E_p$ = 12 MeV presenting an important CN component, the parameter search started with the selection of the CN $J^{\pi}$. The differences in the line shapes and $\gamma$-ray angular distributions for the different $J^{\pi}$  were in most cases marked, such that the choice was evident when comparing with measured data. After selection of the CN state $J^{\pi}$, the branching ratio $W_{l_0}$ of CN state decay orbital angular momenta and the proportion of CN ($\equiv$$W_{CN}$) and direct reaction component (1-$W_{CN}$) were scanned, aiming for a simultaneous good reproduction of the line shapes in all detectors and the $\gamma$-ray angular distribution. Practically, for each $W_{l_0}$, varied in steps in steps of 0.05 - 0.2, the parameter $W_{CN}$ was found in a fit of the angular distribution and the resulting line shapes calculated and compared to the data.  Usually, the line shapes at most or all detector angles and the angular distribution had their best agreement with experiment for very similar parameter values. 

The results of the parameter search to reproduce the  measured line shapes and $\gamma$-ray angular distributions are summarized in table \ref{tabres}. They have been obtained with the potential of  Meigooni et al. \cite{Meigooni85} for the direct reaction component, except at $E_p$ = 5.44, 5.95, 7.0, 7.6 and 8.23 MeV, where slightly better results could be obtained with a potential from the compilation of Perey\&Perey \cite{Perey}. In nearly all cases, the $J^{\pi}$ of the best adjustment corresponds  to a resonance state in $^{13}$N whose energy is within $\Gamma_{tot}$, the total with of the state, to the excitation energy $E_x$ in the CN. At $E_p$ = 9.2 MeV, only the $\gamma$-ray angular distribution is available, however, because of the importance of the cross section peak, the data were included in the present study.

\begin{table} 
\caption{Results of line-shape and $\gamma$-ray angular distribution calculations for the Orsay experiments. $E_p$ and $\Delta E_p$ are the beam energy and energy loss in the target in MeV, respectively, and $E_x$ the excitation energy in the CN.  $J^{\pi}$ and $W_{CN}$ are the spin-parity and proportion of the CN component of the best adjustment.  The probable corresponding state in $^{13}$N, with its excitation energy in MeV (values from \cite{NNDC}), is reported in column 9. $\chi^2$(C) is the result of the least-squares fit of the measured data with the calculated angular distributions and $\chi^2$(L) the result of the corresponding Legendre-polynomial fit. Values in italics for $W_{l_0}$, the proportion of the lowest possible decay angular momentum,  are not constraint.}
\smallskip

\small\noindent\tabcolsep=9pt

\begin{tabular}{lllllllllll}
\\
 $E_p$ ~ &  $\Delta E_p$ ~ & $E_x$ ~ &  $W_{CN}$  &  $J^{\pi}$  ~ & $W_{l_0}$~ & ~ $\chi^2$(C) & ~ $\chi^2$(L) &~ $^{13}$N state  \\
\hline 
5.44 & 0.04 & 6.9 & 0.63 & $\frac{3}{2}^+$ & 0.9  & ~ 0.5 & ~ 0.09 & ~ $\frac{3}{2}^+$, 6.886(8)  \\
5.95 & 0.04 & 7.4 & 0.37 & $\frac{5}{2}^-$  & 0.9 & ~ 1.7 & ~ 0.37 & ~ $\frac{5}{2}^-$, 7.376(9)  \\
6.5 & 0.04 & 7.9 & 0.68 & $\frac{3}{2}^+$ & 0.3  &  ~ 0.18 & ~ 0.002 & ~ $\frac{3}{2}^+$, 7.900 \\
7.0 & 0.04 & 8.4 & 0.58 & $\frac{3}{2}^+$ & 0.2 & ~ 0.21 & ~ 0.01 & ~ $\frac{3}{2}^+$, 7.900 \\
7.6 & 0.04 & 8.9 &  0.61 & $\frac{1}{2}^-$ & {\it 0.5} & ~ 1.00 & ~ 0.46 & ~ $\frac{1}{2}^-$, 8.918(11) \\
8.0 & 0.04 & 9.3 & 0.47 & $\frac{3}{2}^+$ & 0.7 & ~ 0.10  & ~ 0.18 ~ & ~ $\frac{3}{2}^+$, 7.900  \\
8.23 & 0.04 & 9.5 & 0.57 & $\frac{3}{2}^+$ & 0.6 & ~ 2.7 & ~ 0.13 &  ~ $\frac{3}{2}^+$, 7.900  \\
8.4 & 0.04 & 9.7 & 0.47 & $\frac{3}{2}^+$ & 0.8 & ~ 1.6 & ~ 0.02  & ~ $\frac{3}{2}^+$, 7.900  \\
8.6 & 0.16 & 9.8 & 0.26 & $\frac{1}{2}^+$ & 1.0  & ~ 25.1 & ~ 9.5 & ~ ($\frac{1}{2}^+$), 10.25(15) \\
9.0 & 0.15 & 10.2 & 0.41 & $\frac{1}{2}^+$ & 1.0     & ~ 8.3 & ~ 6.1 & ~ ($\frac{1}{2}^+$), 10.25(15)  \\
9.2$^a$ & 0.15 & 10.4 & 0.52 & $\frac{1}{2}^+$ & 1.0   & ~ 7.6 & ~ 7.0 & ~ ($\frac{1}{2}^+$), 10.25(15) \\
9.2$^a$ & 0.15 & 10.4 & 0.55 & $\frac{7}{2}^-$ & 0.0   & ~ 5.7 & ~ 7.0 & ~ $\frac{7}{2}^-$, 10.360  \\
9.6 & 0.15 & 10.7 & 0.54 & $\frac{1}{2}^-$ & {\it 0.5}     & ~ 3.6 & ~ 3.8 & ~ $\frac{1}{2}^-$, 10.833(9)   \\
10.0 & 0.14 & 11.1 & 0.44 & $\frac{5}{2}^+$ & 0.2   & ~ 3.3 &  ~ 3.5 & ~ $\frac{5}{2}^+$, 11.530(12)  \\
10.0 & 0.14 & 11.1 & 0.36 & $\frac{1}{2}^-$ &  {\it 0.5}   & ~ 3.2 & ~ 3.5 & ~ $\frac{1}{2}^-$, 10.833(9)    \\
10.6 & 0.14 & 11.7 & 0.71 & $\frac{3}{2}^+$ & 0.1     & ~ 2.9 & ~ 3.1 & ~ $\frac{3}{2}^+$, 11.740(40)   \\
11.0 & 0.13 & 12.0 & 0.97 & $\frac{7}{2}^-$ & 0.7     & ~ 7.1 & ~ 6.4 & ~ $\frac{7}{2}^-$, 12.130(50)  \\
11.4 & 0.13 & 12.4 & 0.78 & $\frac{7}{2}^-$ & 0.7     & ~ 7.2 & ~ 7.9 & ~ $\frac{7}{2}^-$, 12.130(50)  \\
12.0 & 0.12 & 13.0 & 0.35 & $\frac{7}{2}^-$ & 0.4     & ~ 10.9 & ~ 10.2 &  ~ ( ), 12.937(24)  \\
\hline  
 \end{tabular}
 $^a$ no line shapes available at this energy, results are based on adjustments of the $\gamma$-ray angular distribution only
\label{tabres}
\end{table}

\begin{figure}[htb]
\centering
\includegraphics[scale=0.42]{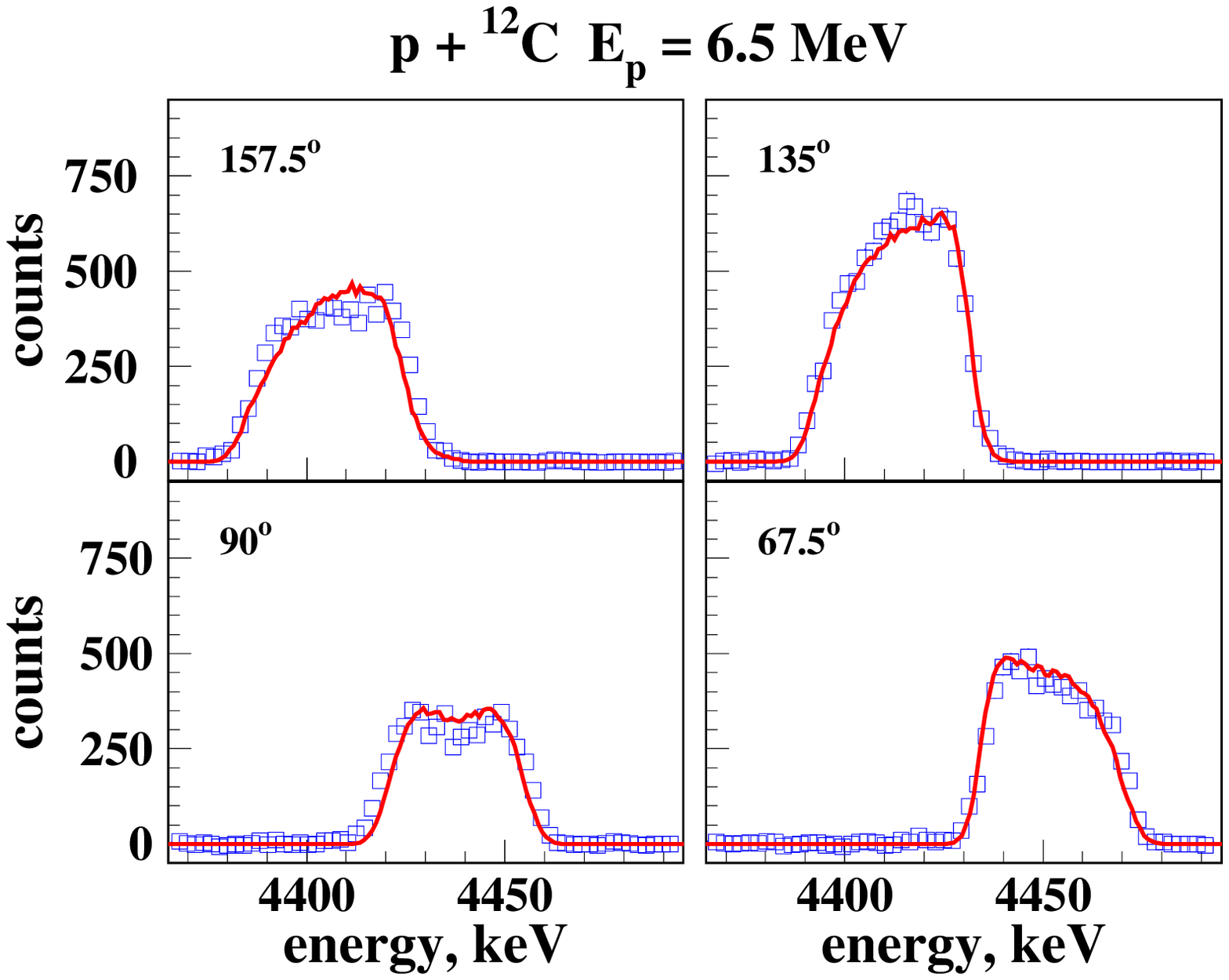} \includegraphics[scale=0.42]{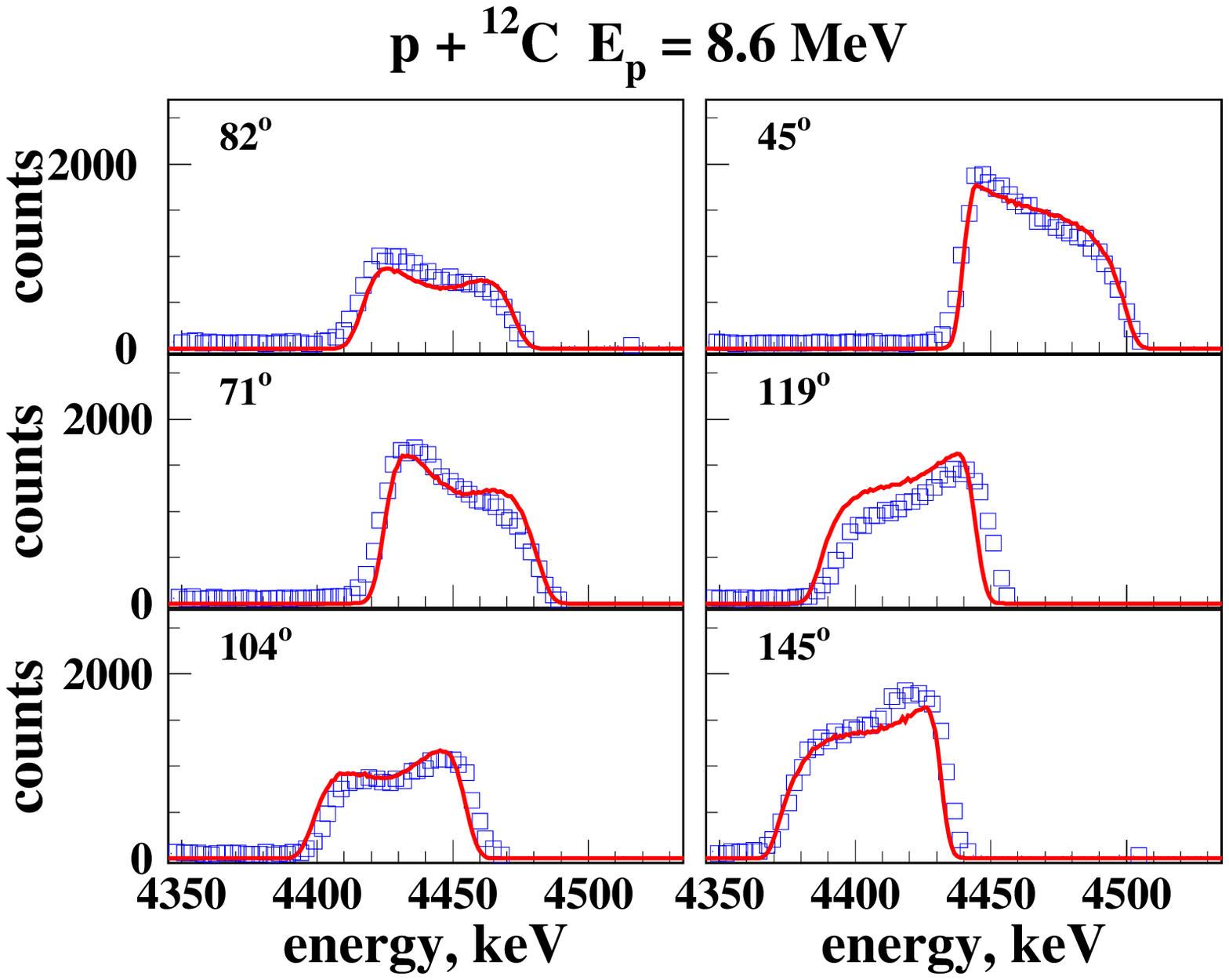} \includegraphics[scale=0.42]{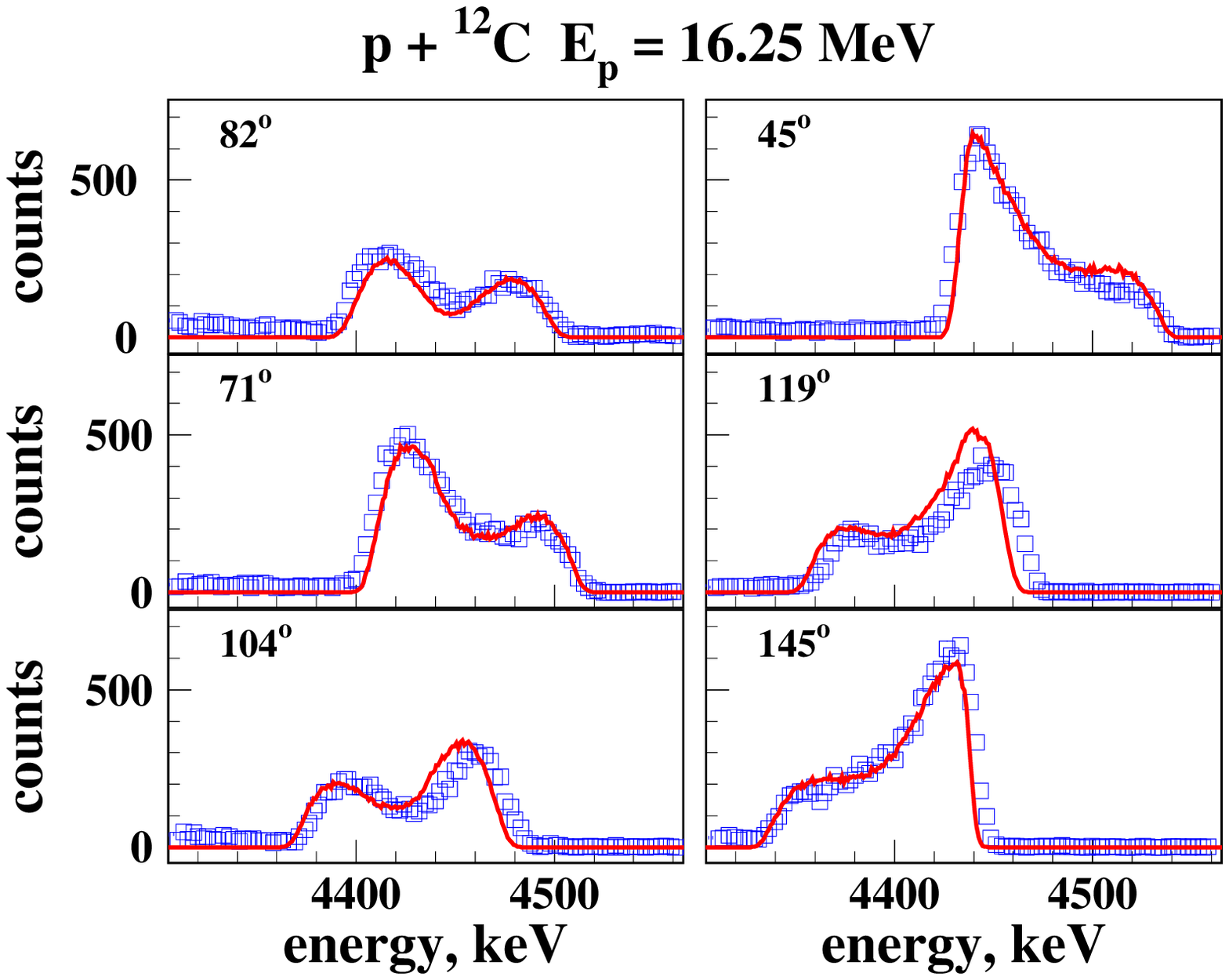} \includegraphics[scale=0.42]{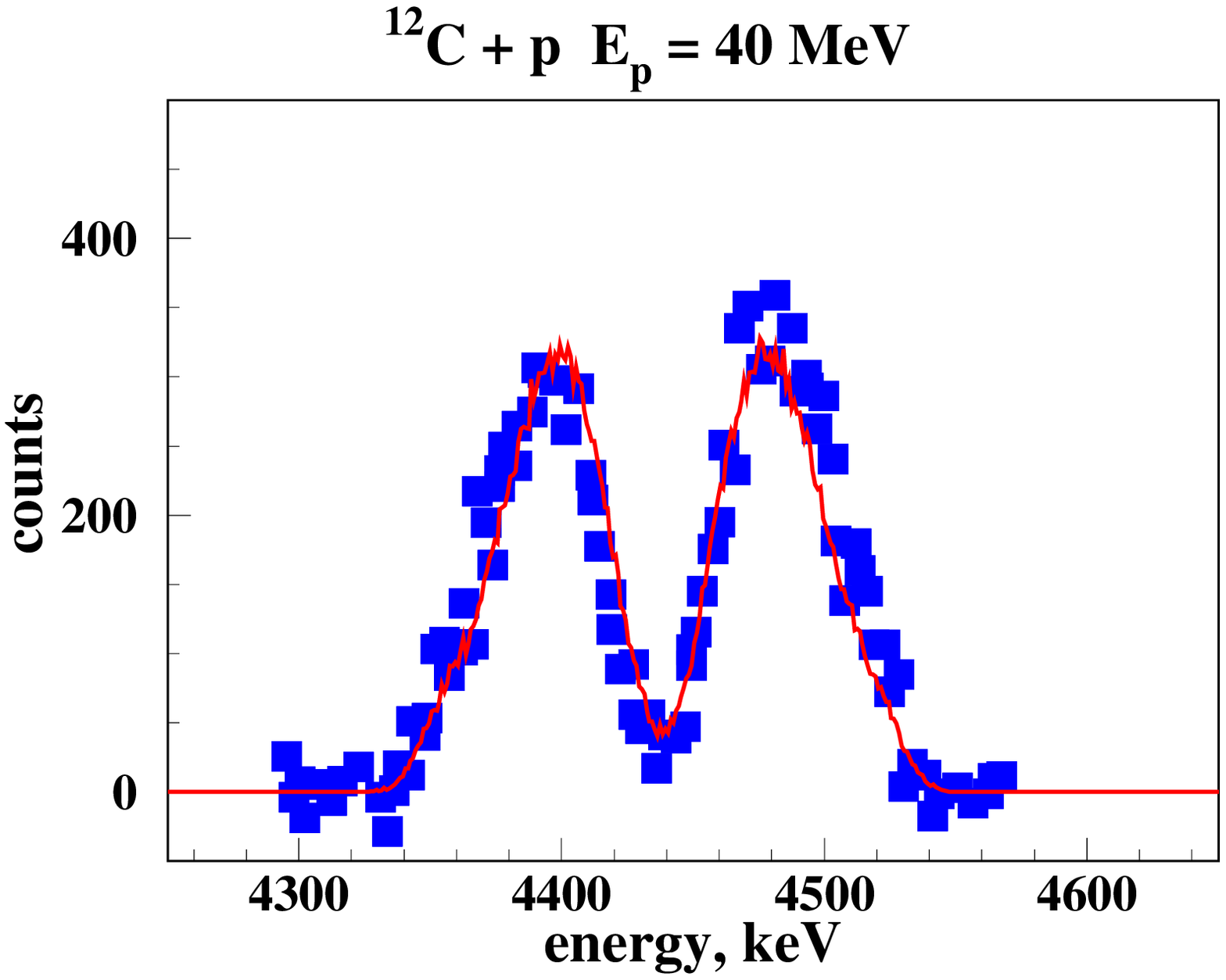} 
\caption[]{Measured shapes of the 4.439-MeV $\gamma$-ray line from proton-inelastic scattering off $^{12}$C in the Orsay experiments  (blue symbols) and results of the line shape  calculation with parameters of table \ref{tabres} (red lines) at $E_p$ = 6.5 MeV (upper left) and $E_p$ = 8.6 MeV (upper right). Examples with dominant direct reaction component at $E_p$ = 16.25 MeV from Orsay and at $E_p$ = 40 MeV from Lang et al. \cite{Lang87} are shown in the lower left and right, respectively. }
\label{figshapes1}
\end{figure}

As expected, there is a sizeable CN component reaching more than 70\% around $E_p$ = 11 MeV.  Remarkably, the adjustments of the $\gamma$-ray angular distributions are very good, resulting, with a few exceptions, in either $\chi^2$ values around 1 or close to those of Legendre-polynomial fits to the data. Line shapes are generally also well described by these calculations, as can be seen in Figure \ref{figshapes1} for some examples.  
 
Data at higher proton energies could be well described with a dominant direct reaction component,  using the potential of Meigooni et al. \cite{Meigooni85} in the coupled-channels calculations. The CN component $W_{CN}$ has been found to be smaller than 20\% for the Orsay data up to $E_p$ = 25 MeV. A typical  example of agreement between calculated and measured line shapes at these energies can be seen  in Figure \ref{figshapes1}. 

The only published line shape at higher energies is from Lang et al. \cite{Lang87}  at a detector angle of 90$^{\circ}$ for a proton beam energy of $E_p$ = 40 MeV. Taking again the potential of Meigooni et al. and 100\% direct reaction component, the calculation reproduces well the measured shape as shown in Figure \ref{figshapes1}. At this energy, nuclear reaction codes like Talys \cite{Talys} predict a non-negligible contribution of the $^{12}$C(p,2p)$^{11}$B reaction populating the 4.445-MeV state. However, simulations based on the Talys calculations predict also a relatively flat line shape between 4.4 and 4.5 MeV for this component. The good reproduction of the measured shape with the inelastic scattering reaction and the deep trough at the nominal energy of 4.438 MeV indicate that the $^{11}$B line has only a very small contribution at this energy. 

\section{Discussion and conclusion}

\begin{figure}[htb]
\centering
\includegraphics[scale=0.35]{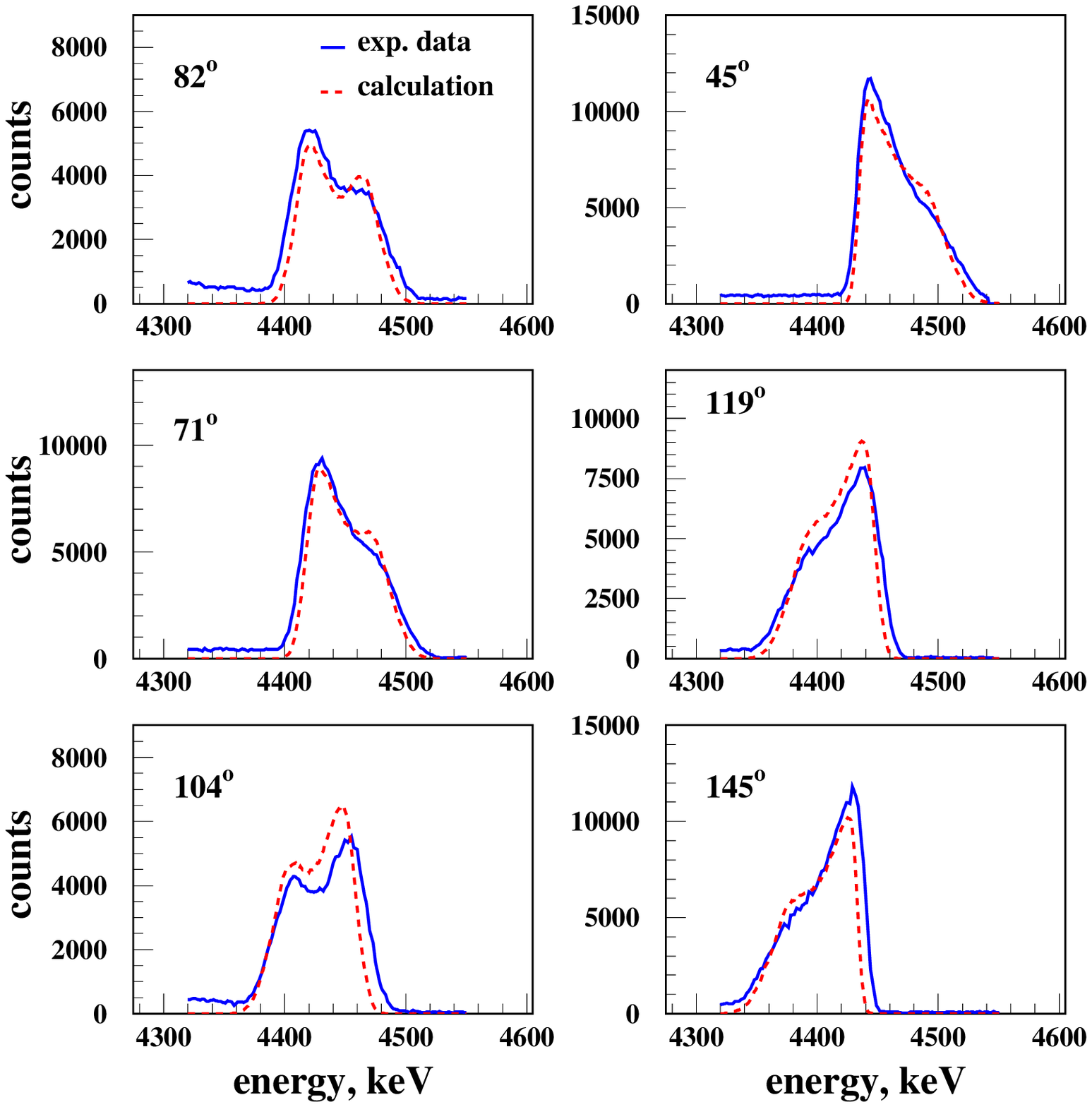} \includegraphics[scale=0.35]{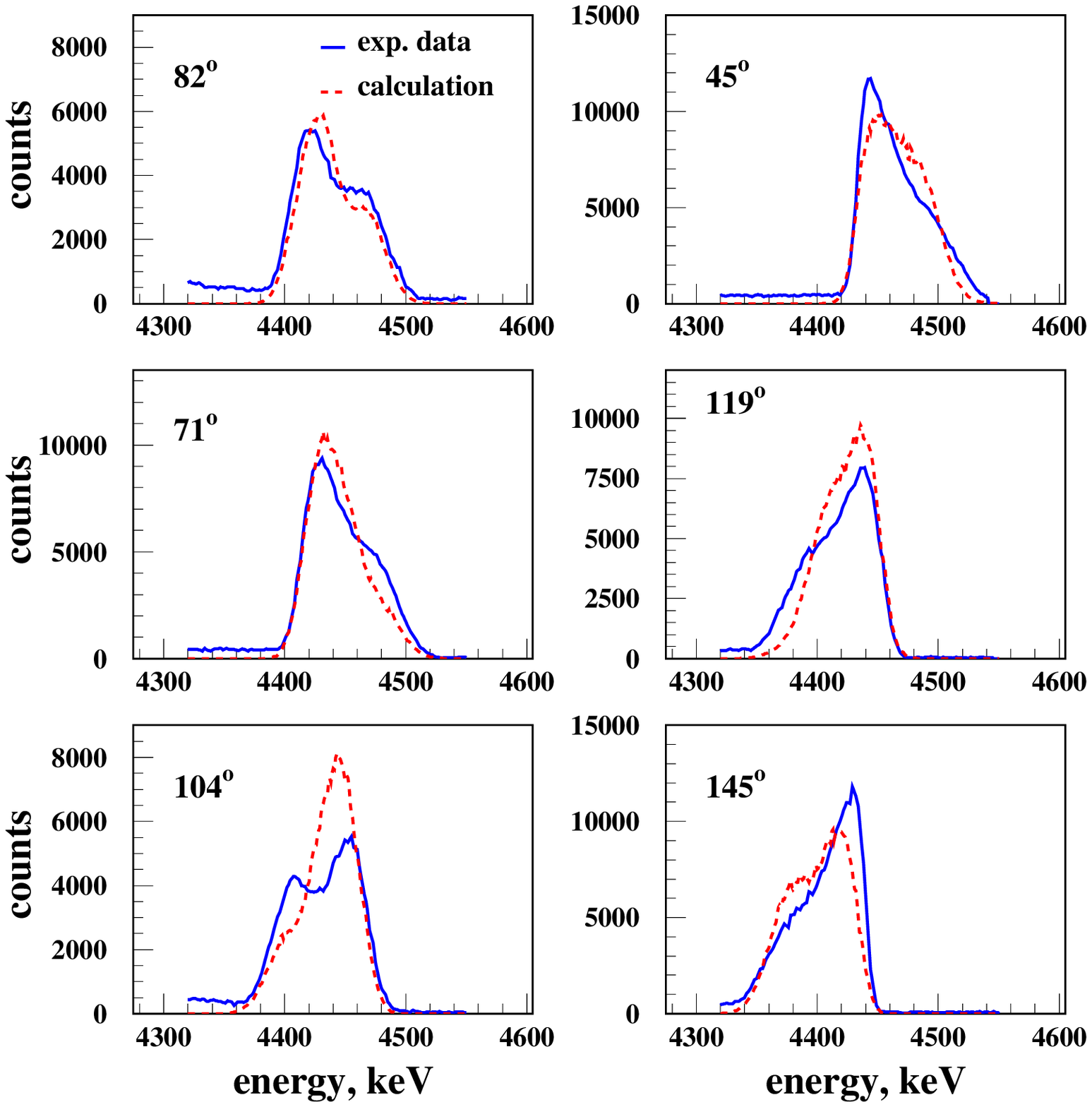}
\caption[]{Weighted sums of measured shapes of the 4.439-MeV $\gamma$-ray line from proton-inelastic scattering off $^{12}$C in the Orsay-1997 experiment  (blue lines) and calculated line shapes (dashed red lines) from the present study (left) and the previous study (right) of Ref. \cite{lshape}.  }
\label{lshape97}
\end{figure}

The most striking improvement in the actual study over previous calculations is the much better reproduction of the measured line shapes in the region of prominent CN resonances and the simultaneous excellent description of $\gamma$-ray angular distributions. This is illustrated in Figure \ref{lshape97}, where weighted sums $P(E_p)$ in the range $E_p$ =  8.52 - 19.75 MeV, with $P(E_p)$ following a typical  interaction probability as encountered in solar-flare $\gamma$-ray emission, of calculated and measured line shapes of the Orsay-1997 experiment \cite{torion} are shown. Furthermore, data from the Orsay-2002 experiment \cite{t2002} that extend the energy range of available line shapes down to threshold and up to $E_p$ = 25 MeV are also well reproduced. 

Taken that the available line shapes and $\gamma$-ray angular distributions cover neatly the explored energy range, with data close to every prominant CN resonance visible in the inelastic cross section excitation function, see Figure \ref{Excf44}, the present calculations can reliably predict the $\gamma$-ray emission in p + $^{12}$C reaction up to $E_p$ = 25 MeV. At energies above $E_p$ = 12 MeV, the data are well described by the direct reaction component from coupled-channels calculations with the potential of Meigooni et al. \cite{Meigooni85}. This energy-dependent nuclear reaction potential has been derived for inelastic nucleon scattering off $^{12}$C in the range up to $E_p$ = 100 MeV, offering the possibility of extrapolating the results of the present study to higher energies, whereby good agreement with a published line shape at $E_p$ = 40 MeV was already obtained.

In impulsive solar flares, the 4.438-MeV line is mainly produced in reactions of energetic protons and $\alpha$ particles with $^{12}$C and $^{16}$O, and in reverse kinematics. Taking typical incident energetic particle spectra and abundances in the solar atmosphere, slightly more than half of the 4.438-MeV   
$\gamma$ rays is produced by proton inelastic scattering off $^{12}$C in the range up to $E_p$ = 100 MeV, and more than 90\% of this component is from reactions below $E_p$ = 25 MeV.  In Ref. \cite{lshape} it was shown that the 4.438-MeV $\gamma$-ray emission in proton and $\alpha$-particle reactions with $^{16}$O  can be simply parametrized, which leaves only the minor contribution of $\alpha$-particle inelastic scattering off $^{12}$C without comprehensive line-shape studies. The present study, that  covers thus practically completely p + $^{12}$C reactions should therefore allow accurate line-shape predictions for impulsive solar flares.

Similar to solar flares, the 4.438-MeV line in human tissue during proton radiotherapy is essentially produced by inelastic scattering off $^{12}$C and reactions with $^{16}$O. Incident proton energies are in the range of about 60 MeV to 200 MeV. At an incident proton energy of $E_p$ = 68 MeV, typical for eye cancer treatment, slightly more than half of the $\gamma$-ray emission in reactions with $^{12}$C is produced for proton energies above $E_p$ = 25 MeV, and about 80\% in reactions with $^{16}$O. The  calculations are thus strongly based on extrapolations, but line shapes and angular distributions should nonetheless be quite accurate, as explained above. 

This may not hold for proton energies well above $E_p$ = 100 MeV, where other nucler reaction potentials may be needed for the inelastic scattering off $^{12}$C and cross sections of the 4.438-MeV line emission in proton reactions with  $^{16}$O have only been measured up to $E_p$ = 50 MeV. Fortunately, new experimental data for p + $^{12}$C and p + $^{16}$O reactions up to $E_p$ = 200 MeV from the iThemba LABS cyclotron should soon be available. First results are encouraging \cite{Yahia}, such that more accurate predictions in the cited applications can be possible in the near future.

\section*{Acknowledgements}
This study wouldn't have been possible without important help of my collegues I. Deloncle, V. Tatischeff, J.-P. Thibaud at CSNSM, N. de S\'er\'eville at IPNO, F. de Oliveira at GANIL, H. Benhabiles at Univ. Boumerdes, A, Belhout, S. Ouichaoui and W. Yahia-Cherif at USTHB Alger and the commitment of the staff at the Orsay tandem accelerator during the experiments.

\end{document}